\documentclass{svjour3}                      
\smartqed  
\usepackage{graphicx}
\usepackage{lineno,hyperref}

\usepackage{graphics,color}
\usepackage{amssymb}
\usepackage{booktabs}
\usepackage{mathtools}
\usepackage{algorithmic}
\usepackage[ruled,linesnumbered,boxed,commentsnumbered,lined]{algorithm2e}
\usepackage{adjustbox}
\usepackage{multicol}
\usepackage{url}
\usepackage{lscape}
\usepackage{booktabs}
\usepackage{multirow}
\usepackage{pifont} 
\usepackage{indentfirst}
\usepackage{parskip}
\usepackage{placeins}
\usepackage{float}

\begin{document}
\title{Modeling User Behaviour in Research Paper Recommendation System}

%


\author{Arpita Chaudhuri$^{1}$ \and 
        Debasis Samanta$^{1}$ \and Monalisa Sarma$^{1}$
}

\authorrunning{Chaudhuri et al.} 

\institute{$^{1}$ A. Chaudhuri, D. Samanta, M. Sarma \at
              Indian Institute of Technology Kharagpur, Kharagpur, India \\
              \email{dsamanta@iitkgp.ac.in}           
                              }

\date{Received: date / Accepted: date}

\maketitle

\begin{abstract}
  User intention which often changes dynamically is considered to be an important factor for modeling users  in the design of recommendation systems. Recent studies are starting to focus on predicting user intention (what users want) beyond user preference (what users like). In this work, a user intention model is proposed based on deep sequential topic analysis. The model predicts a user’s intention in terms of topic of interest. The Hybrid Topic Model (HTM)  comprising Latent Dirichlet Allocation (LDA) and Word2Vec is proposed to derive the topic of interest of users and the history of preferences.  HTM finds the true topics of papers estimating word-topic distribution which includes syntactic and semantic correlations among words. Next, to model user intention, a Long Short Term Memory (LSTM) based sequential deep learning model is proposed. This model takes into account of temporal context, namely the time difference between clicks of two consecutive papers seen by a user. Extensive experiments with the real-world research paper dataset indicate that the proposed approach significantly outperforms the state of the art methods. Further, the proposed approach introduces a new road map to model a user activity suitable for the design of a research paper recommendation system.
  
\end{abstract}

\section{Introduction}
\label{intro}
With the growing repository of information in the Internet, the existing search engines are failed to satisfy user's demand \cite{1} efficiently. One reason is that they do not consider user profile into account. In fact, for a given search query, a search engine provides the same information to all users, although individual users may have  their own preferences. This drawback  necessitates a personalized information system whose goal would be to satisfy a user's preference based on her own need. Of late, user's personification becomes one of the current research interest and becoming popular in several domains, such as artificial intelligence, data mining, information science, etc. \cite{2}. Another, the most sought application is the recommendation system \cite{3,21}. A recommendation system, in general, recommends a list of items out of a large pool of items based on user's interest(s). More precisely, it should generate distinct information given a specific search query, but issued by different users. Thus, a good user profile plays an important role for an interactive model and enhance the performance of the system. Personification  through user profile is adapted in the design of recommendation systems in different domains,  for example, movies, songs, books, videos, web pages, e-learning, articles, etc. \cite{4,22}. Among all these, recommendation of articles, that is, research paper recommendation system deserves more attention. 

Traditional studies of research paper recommendation systems primarily focus on estimating intrinsic user preferences, which is consistent as time passes and also are not perfectly represented. For example, the typical TF-IDF \cite{5,14} model represents user and item using vector of words. Further, ``topic of interest" \cite{9,18,19,17,20} or `keyphrase" \cite{8,10}, extracted from metadata such as title, abstract, keyword, whole text of user's tagged or published papers are also used to represent user preference. Another approach of representing user preference is collecting information explicitly \cite{10,14}.\\

\indent Although the above-mentioned approaches represent user profiles from several views, they are not free from issues. The ``bag of words" model leads to process a huge information and can not detect syntactical similarity among words. The ``keyphrase" model uses the meaningful contextual information to define user profile, it also involves the processing of a large amount of information. For the identification of topics from papers, the existing solutions merely consider correlation of words. This may lead to inaccurate identification of topics. User profile based on external information gives an overhead to users and usually suffers from sparsity problem due to lack of interest in providing information. Moreover, the user preference remains unchanged while making recommendation in different context. Another issue is that in real-life applications user intention should be beyond user preference and better be governed by past interactions. Further, a paper belong to a preferred topic of a user may not be liked next time by that user. Hence, it is required to concentrate on what users want (user intention) after what users like (user preference). In capturing user recent intention, sequential algorithms \cite{39,40,41,45} are good and recently have gained attention among researchers. The sequential prediction model predicts user's next behaviour from recent interaction. The dynamic user intention modeling has been justified well in many text-based recommendation systems, such as web page recommendation \cite{23,24,25}, news article recommendation \cite{26}, e-learning \cite{46} etc. Nevertheless, for a research paper recommendation system, an improved approach is yet to be reported.

\indent To address the aforementioned limitations, this work finalizes the following objectives.
\begin{itemize}
    \item To propose a topic model that overcomes the drawbacks of the existing topic modeling approaches.
    \item To model user intention from previous interactions with topic of interest considering time stamp as a temporal context.
    \item To check the performance of a research paper recommendation system considering the proposed user intention modeling.
    \end{itemize}
In realization of the above objectives, a comprehensive model of ``user preference learning" and ``user intention prediction" is proposed. In order to learn the user preference, this work proposes a hybrid topic model. The hybrid topic model combines Latent Dirichlet Allocation (LDA) and Word2Vec to study the  probabilistic distribution of words over topics of papers followed by contextual relationship of words. Afterwards, an algorithm is proposed to decide the true topic of papers. Next, the user preferences are collected from the user's log file in terms of topic of interest. The preferable topics are extracted from the clicked papers only. Further, a sequential model comprising with Long Short Term memory (LSTM) is used to model user intention from the historical sequence of topic of interest. The  model utilize time stamp as a temporal context feature. This framework enables to model dynamic behaviour of users considering long and short term analyses.\\  

\indent The main research contributions are highlighted as follows;
\begin{itemize}
    \item This work proposes a hybrid topic model to identify topic of a paper.
    \item The proposed hybrid topic model alleviates all the drawbacks of existing topic models.
    \item A user intention prediction model is presented which able to capture users dynamic interest or demand at a particular moment.
    \item The proposed topic model is applied to the dataset of two thousands records collected from the Scopus repository and demonstrate the effectiveness of the proposed topic model with comparing to the existing model. Also, the effectiveness of user intention model is proved by comparing with the baseline model. 
\end{itemize}
The rest of the paper is organized as follows. Section \ref{sec:rw} presents the related work associated with the summarizing of the characteristics of a user profile. In Section \ref{sec:pa}, the proposed user preference learning and user intention prediction are discussed. Section \ref{sec:exp} provides the experiment and analysis of results. Next, threats to validity of the proposed approach are discussed in Section \ref{sec:tv}. Finally, Section \ref{sec:con} concludes the paper.

\section{Related work}
\label{sec:rw}
The state of the art research paper recommendation systems mostly emphasize on the preference learning of users. This section outlines some of the existing works related to the different techniques of modeling user behaviour.\\
\indent Sugiyama et al. \cite{5} proposed a paper recommendation system, where  a user profile is represented using a feature vector comprising unique terms obtained from a researcher's past published papers. Each term is defined by term frequency-inverse document frequency. Subsequently, papers were recommended using similarity matching between feature vectors of candidate papers and user preferences.\\
\indent Dhanda and Verma \cite{7} presented a recommender system based on incremental dataset. The system collected publication date, publishing authority as a preference from user's liked paper. Further this user preferences were used to generate output papers.\\
\indent Hong et al. \cite{9,18} defined user profile by the topic given by a user. It is updated on a new topic provided by the user. Then cosine similarity was used to find related papers for recommendation.\\ 
\indent A deep learning based research paper recommendation system proposed by Hassan \cite{20} to create user profile based on implicit and explicit feedback of a user. In the first step, explicitly they collected user's preferred articles and extracted title and abstract to create a vector of topics. Next, user's short term interest is collected implicitly from the transaction log. Finally, papers were recommended using cosine similarity between the candidate papers and papers that were stored in user profile.\\
\indent Alshaikh \cite{19} introduced an interested way of representing user profile. In his content based research paper recommendation system, a Dynamic Normalized Tree of Concepts (DNTC) is used to build user profile. The tree maintains the parent-child relationship between concepts following ontology. From the user's each reading paper, top N concepts are retrieved. further, a tree with weight $0$ is build to explore the semantic relation between concepts. The tree is normalized by the number of reading papers of a user and dynamically updated based on the time sequence of user's log data. If a user reads new paper in new time, the weight of the tree is recalculated to update. The DNTC represents user's short term interest. Afterwards, a distance measure was used to find the most similar papers related to user short term interest. \\ 
\indent Alshaikh \cite{17}  incorporated long term and short term interest of users to present his profile. Here, a dynamic sliding window was used with DNTC to reflect short term interest of a user. the length of the sliding window depended on the latest papers read by the user. The concept of all papers read by the user considered as a long term preference. After that, using a tree distance measure papers are recommended that matches with their long term interest. \\ 
\indent Ferrara et al. \cite{8} proposed a content based recommendation approach where user profile was constructed from the tagged paper of users. From each paper, the weighted keyphrases such as uni-gram, bi-gram, and tri-gram are calculated. These three different list of keyphrases created three different user profile. The weight of each keyphrase was multiplied by inverse document frequency of associated keyphrase. In the final step, similarity matching was done between candidate papers and user profile\\
\indent Sun et al. \cite{11} suggested a hybrid article recommendation in social networks. The metadata of published papers in social network for example title, abstract, published journal, year were used to define profile. The authors utilized three connectivity graphs to find the relation between user-user, user-article and user-keywords. These three relations were represented by three matrices. Finally, Random Walk Restart (RWR) method was employed to find recommended articles.  \\
\indent Gautam et al. \cite{10} collected personal information of users explicitly. They used user given tags or keywords associated with their academic data to build profile. Further, cosine similarity was used to recommend papers.\\ 
\indent Bulut et al. \cite{12,13} provided a feature based user profile. To generate profile, the authors considered a user's past publication. All the required metadata such as title, year, author, abstract, and keyword of each article was extracted and merged together in a profile. In this work, also cosine similarity was utilized to find similar interested papers.\\ 
\indent Wang et al. \cite{14} proposed a hybrid article recommendation system where, for the content based part, they assumed user preference from user given tag and title of the article read by a user. Further, they represent user preference in terms of weighted bag of words to create user profile. Finally, recommended articles were found out my matching similarity between article profile and researcher profile.

The characteristics of user profiles, modeled in different article recommendation systems are summarized in Table \ref{tab:table1}. From the survey, it can be observed that all the existing works considered user preference only that is what user likes to decide their relevant papers. Whereas, capturing change of preferences means what user wants is very important for accurate recommendation. It is completely missing in the existing approaches. 

\begin{table*}[b]
\centering
\caption{Characteristics of user modelling in different schemes.}
\label{tab:table1}
\resizebox{12.5cm}{!}{
\begin{tabular}{cccccc}
\hline
& & \\
\multicolumn{1}{c}{Sl. No.} & \multicolumn{1}{c}{Schemes}                            & \multicolumn{1}{c}{Modelling technique} & \multicolumn{1}{c}{SPL} &  \multicolumn{1}{c}{Intention prediction} &   \multicolumn{1}{c}{LSTPA}                                                       \\  & & \\ \hline 1                             & \cite{5}              & Bag of words    &   past publication   &   No    & No                                                                                  \\ & & \\
2                             & \cite{7}                & Publishing date, \\ & &  publishing authority    &   Explicit feedback & No  & No                                                      \\ & & \\
3                             & \cite{9,18}                  & Topics of interest & Explicitly from user & No &  No
\\ && \\
4                             & \cite{20}                   & Topics    & implicit and explicit feedback  & No & No                     \\ && \\
5                             & \cite{19}                  & Tree of concepts  &   past publication & No & No

\\ & & \\
6                             & \cite{17}               & Tree of concepts  &   past publication & yes  &  Yes                                                     \\ & & \\
7                             & \cite{8}             & Keyphrase   &   Tagged paper  & No   &  No                                                   \\ & & \\
8                             & \cite{11}                  & Metadata of papers   &  social network   & No   &    No                                             \\ & & \\
9                             & \cite{10}               & Explicit information, tagged keyword  &  Explicit feedback  &  No  &  No  
\\ & & \\
10                            & \cite{12,13}            & Features of papers  &  Past publication  & No  & No                                                                                \\ & & \\
11                            & \cite{14}         & Bag of words  &   user given tag   &   No  &   No                                                      
\\ & & \\
\hline
\end{tabular}}
\caption*{SPL: Source for Preference Learning, LSTPA: Long-Short Term Preference Analysis}
\end{table*}

\section{Proposed approach}
\label{sec:pa}
For the recommendation system, user behavior monitoring is a crucial task. In this context, proposed approach describes the strategy of extracting user's preference and predicting user intention for accurate recommendation. For a given user, the papers that she clicked from a recommendation list are considered as relevant papers for building user profile. These relevant papers are traced from user browsing activities. Suppose, a user $u_i$ has clicked a set of papers $P = {p_1, p_2, \cdots, p_n}$. Now, the primary goal is to extract the preference of user $u_i$ in terms of topic of interest from each relevant paper $p_i$. Further, consider the historical preference of the user $u_i$ is $H^{u_i} = [(tp_1^{u_i}, t_1^{u_i}), (tp_2^{u_i}, t_2^{u_i}), \cdots, (tp_n^{u_i}, t_n^{u_i})]$, where, $(tp_i^{u_i}, t_i^{u_i})$ is the $i$-th topic of interest of user $u_i$ at time $t_i$. Hence, the second target is to predict the intention of user $u_i$ at time $t_{n+1}$.

\begin{figure}[b]
\centering
\includegraphics[width=5.1in]{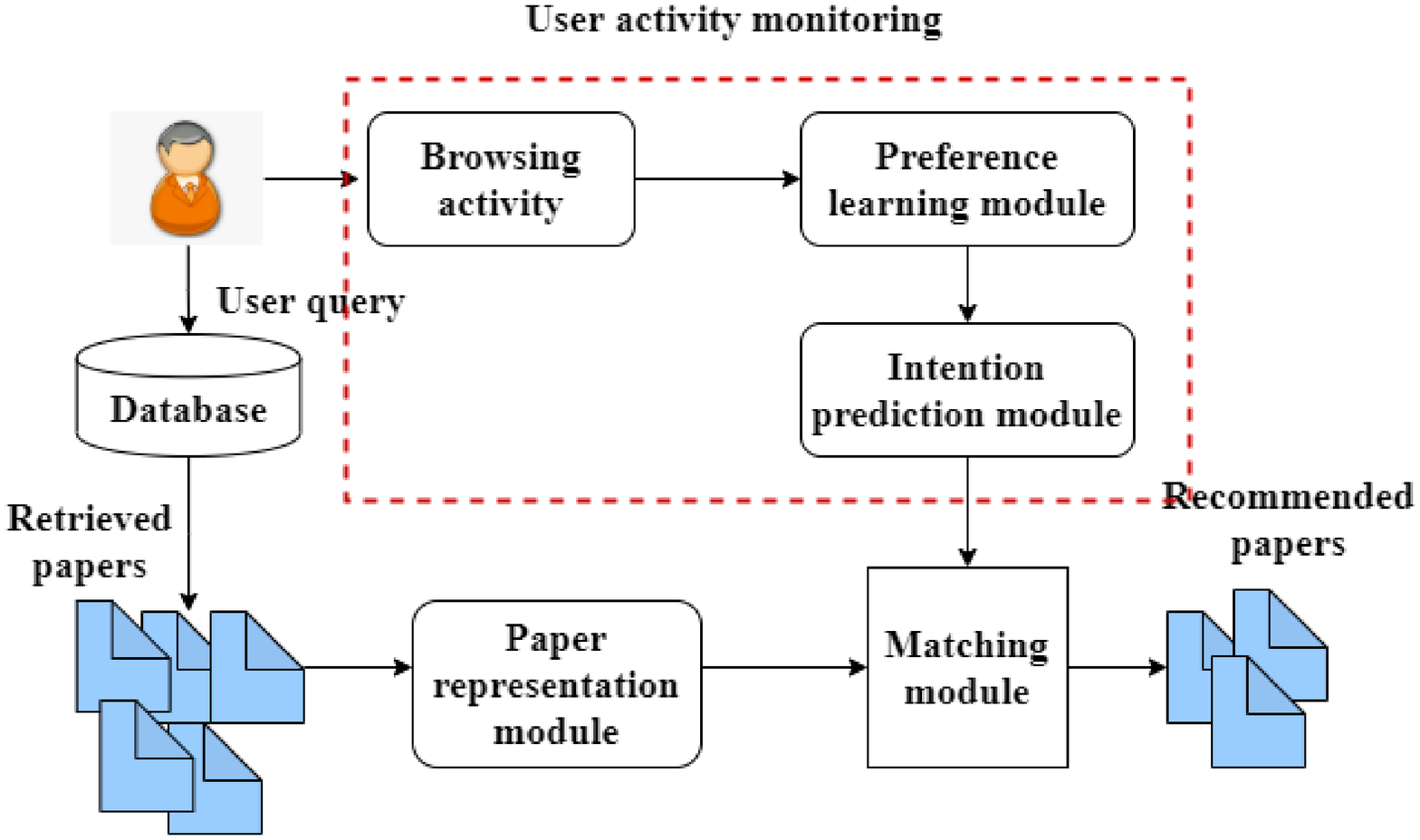}
\caption{Overview of the proposed approach }
\label{Figure:fig1}
\end{figure}

\indent \textbf{Overview of the proposed approach} The proposed approached is categorised in two phases:  (1) User preference learning and (2) User intention prediction. In the first phase, it explains the procedure to learn user preference implicitly. Further, the proposed approach presents LSTM based sequential model to predict user intention. An overview of the proposed approach is shown in Figure \ref{Figure:fig1}.

\subsection{User preference learning} User preference plays an important role to generate user profile in a recommendation system. An accurate user profile easily deal the personalized recommendation. There are two methods to capture user preference (1) Explicit and (2) Implicit. In explicit method, users have to provide their personal details or preferences by themselves and this creates burden to users. Hence, the explicit capturing is not a well accepted method and users may loose their interest shortly. Whereas, implicit method involves capturing user's preference without any intervention of users. In this regard, the proposed method  collect user preferences implicitly considering user's clicked data. In the research paper recommendation system, the clicked data refers a set of papers. A paper contains several features. The proposed method considers a latent features of papers that is ``topics" as preference of users. In other words, the proposed approach tries to capture a user's preference in terms of topics of interests. In order to extract the topic of papers, a hybrid model is proposed. The hybrid topic model comprises with Latent Dirichlet Allocation (LDA) \cite{27} and Word2Vec \cite{28} method. For calculating the probability distributions of words of a paper among predefined several number of topics, LDA, a topic modelling techniques is considered. It may be noted that, LDA does not consider the contextual correlation among words and thus fails to predict true topics of words. Word-Word correlation measure is necessary to justify the true identification of topics of a paper. In this context, a word embedding technique Word2Vec is taken into account. Word2Vec assist to group semantically and syntactically similar words under a particular topic. This work proposes an approach how to combine LDA and Word2Vec to obtain perfect word-topic distribution.   

\begin{figure}[!htbp]
\centering
\includegraphics[scale=0.5900, angle=0]{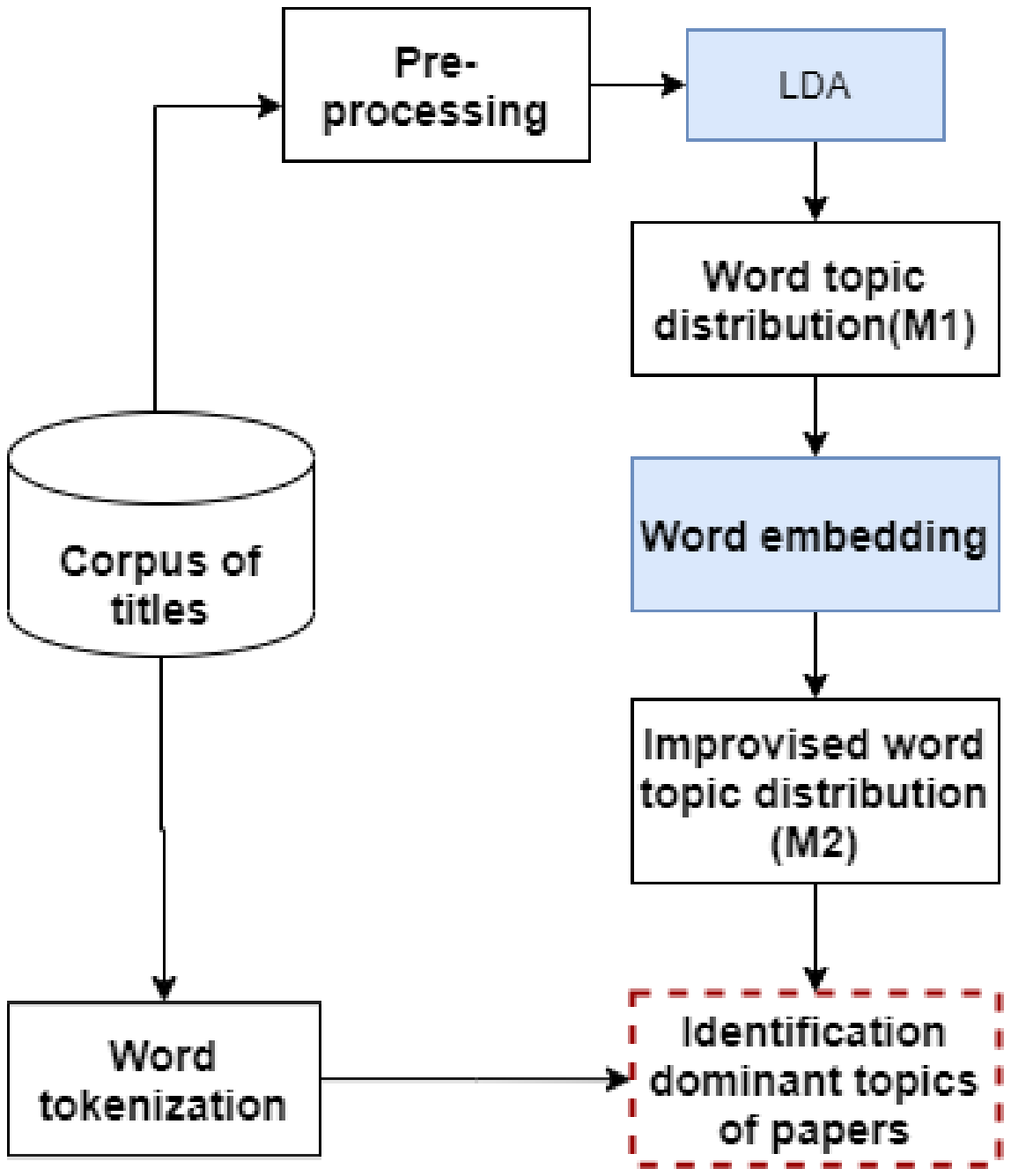}
\caption{Framework of the proposed topic prediction method. }
\label{Figure:figure1}
\end{figure}

Further, the topic of a paper is decided from the maximum likelihood of words among topics. The necessary steps involved in the proposed approach to predict the topic of papers is shown in Fig. \ref{Figure:figure1}.
\subsubsection{Pre-processing} For applying LDA and Word2Vec both, corpus needs to be pre-processed. In this process, a database, say $D$ comprising the title of all papers is considered. Further, the dataset $D$ is pre-processed to remove stop words, punctuation, numbers and so on. Along with this, words tokenization, lemmitization and stemming has been done. So, each title is tokenized into words and it will be the input for next step.

\subsubsection{Calculation of word-topic distribution using LDA} The first step to identify the topic of a paper is calculating the distribution of words over several topics. Here, LDA, an unsupervised machine learning technique is used to quantify the probability of a words to be in a specific topic. To quantify the probability distribution of words among topics using LDA, a ``Bag-Of-Word" (BOW) model is required to estimate. Hence, in the next step, with the tokenized words generated in the pre-process step, a dictionary with unique words is constructed. Then, $D$ is represented as ``Bag-Of-Word" (BOW) model. Now, The LDA is applied to BOW model to calculate probability distribution of words into $k$ number of topics. Here, $k$ is predetermined. The best value of $k$ is decided by fitting LDA several times with different $k$ values and calculating topic coherence score \cite{29} every time. Finally, the value of $k$ for which highest topic coherence score is obtained, is taken as a best value of $k$. Initially, for each title $t_i$, each word $w_i \in t_i$ is assigned randomly to one of $k$ topics and formed a matrix $M_1$. Suppose, The dataset $D$ contains $m$ number of titles. Each title $t_i$ contains $n$ number of words and the total number of words in vocabulary is $v$. Now, for each word in $M_1$, its topic assignment is updated based on two predictions (1) $P(topic(tp_j) \mid title(t_i))$, where $(i = 1, 2,\cdots, m)$ and $(j = 1, 2, \cdots, k)$. This distribution figures out the number of words of given title $t_i$ belong to the topic $tp_j$ and (2) $P(word (w_l) \mid topic (tp_j))$, where,  $(l = 1, 2,\cdots, n)$ and $(j = 1, 2, \cdots, k)$. It calculates the number of titles that indicates the number of papers are in topic $tp_k$ because of the word $w_l$. The steps to calculate above two predictions and the updating of probability is explained below. 
\begin{enumerate}
    \item In the first step of LDA, each title is modeled with Poisson distribution \cite{30} of words.
    \item In second, for each topic the proportion  $p(word(w_l) \mid topic (tp_j))$ which is denoted by the random variable, say, $\overrightarrow{\tau_k}$ is estimated. It is a Dirichlet distribution \cite{30} of the words in topic $tp_j$, parameterized by $\overrightarrow{\beta}$. $\overrightarrow{\beta}$ is $v$ dimensional vector of positive real numbers with sum upto one. The posterior estimation of $\tau_{xy}$ that means probability of word $x$ for topic $y$ can be estimated as;
    \begin{equation}
       \tau_{xy} = \frac{M^{wtp}_{xy} + \beta}{\sum_{w=1}^{v}M^{wtp}_{wy} + v_{\beta}} 
    \end{equation}
    Where, $M^{wtp}$ is the matrix of word-topic count.
    \item The third step samples another random variable, say, $\overrightarrow{\vartheta_m}$ which represents the probability $p(topic(tp_j) \mid title(t_i))$, where $(i = 1, 2,\cdots, m)$ and $(j = 1, 2, \cdots, k)$. It is another Dirichlet distribution of topics for each title $t_i$ and parameterized by $\overrightarrow{\alpha}$. $\overrightarrow{\alpha}$ is a $k$ dimensional vector of positive real numbers with sum upto one. The posterior probability of topic $y$ in title $q$ that is $\vartheta_{yq}$ is computed using following equation;
    \begin{equation}
       \vartheta_{yq} = \frac{M^{ttp}_{yq} + \alpha}{\sum_{t=1}^{k}M^{ttp}_{yq} + k_{\alpha}} 
    \end{equation}
    Where, $M^{ttp}$ is the matrix of title-topic count.
    \item Now, for each title $t_i$, identify the topic of each word $w_l \in t_i$ based on multinomial distribution \cite{30} given $\overrightarrow{\vartheta_m}$. In addition, the other words which are in the same topic are identified from the word distribution $\overrightarrow{\tau_k}$. This process is repeated for all the papers to improve the topic assignment of words. This reassignment of the probability of a word to a topic can be expressed mathematically as follows;
    \begin{equation}
        P(z_i = j \mid z_{-i}, {w_i}, t_i) = \frac{M^{wtp}_{{w_i}j} + \beta}{\sum_{w=1}^{v}M^{wtp}_{wj} + v_{\beta}} \times \frac{M^{ptp}_{{t_i}j} + \alpha}{\sum_{t=1}^{k}M^{ttp}_{{t_i}tp} + k_{\alpha}} 
    \end{equation}
    Where, $P(z_i=j)$ is the probability that the topic $j$ is assigned to the $i$-$th$ word, $z_{-i}$ represents the topic assignment of other words, $t_i$ is the title of a paper contains the $i$-$th$ word, 
    \end{enumerate}
Top 10 words under each topic which is obtained from LDA technique is shown in Figure \ref{Figure:figure2}.
\begin{figure}[!tbp]
\centering
\includegraphics[width=5.5in]{lda_out.eps}
\caption{Top words generated using LDA. }
\label{Figure:figure2}
\end{figure}

\subsubsection{Improvised word-topic distribution using word embedding} Though LDA reduce the dimensionality of text corpus and represents papers in low dimensional vector, it has some drawbacks too. A ``bag-of-word" model is used to represent papers in LDA. Hence, LDA suffers the sparsity problem for representing papers. In addition, it does not produce good result for small training data. Besides this common problems, another important issue is not considering semantic correlation of words during distribution of words into topics. To achieve better word-topic distribution, it is necessary to consider the context of words. In this regard, word embedding is most promising technique. In order to embed words by semantic and syntactic relations, a skip-gram model, variant of Word2vec model is chosen. The Word2vec model is a word embedding technique which represents words of a large text in n-dimensional vector space. It follows the distributional hypothesis \cite{31} that words comes in the similar context have similar meaning. There is a two techniques to train the Word2Vec model: (1) Continuous Bag Of Word (CBOW) and (2) Skip-gram. Skip-gram model is preferable over CBOW model where training corpus is not so large. Hence, in this work skip-gram model is considered. 

Word2Vec forms a vocabulary of unique words considering the tokenized words, obtained in the pre-processing steps from the database $D$. Further, a $n$-dimensional feature space or vector space is created where, each unique word of vocabulary is assigned to a corresponding vector in vector space. Along with this, a training sample with context-target word pairs is prepared depending upon context window size. The context window is a very important hyper-parameter \cite{32} to determine the number of contextual neighbours of the target word while estimating the vector representation. This work has selected the context window size is $6$. In the proposed method, top $2$ words are extracted from the matrix $M_1$ of the previous step. For each word, $6$ context words are found using Skip-gram model.

A Skip-gram model is a fully connected neural network which is constructed with a input layer, one hidden layer and output layer. 
\begin{figure}[!htbp]
\centering
\includegraphics[scale=0.50000, angle=0]{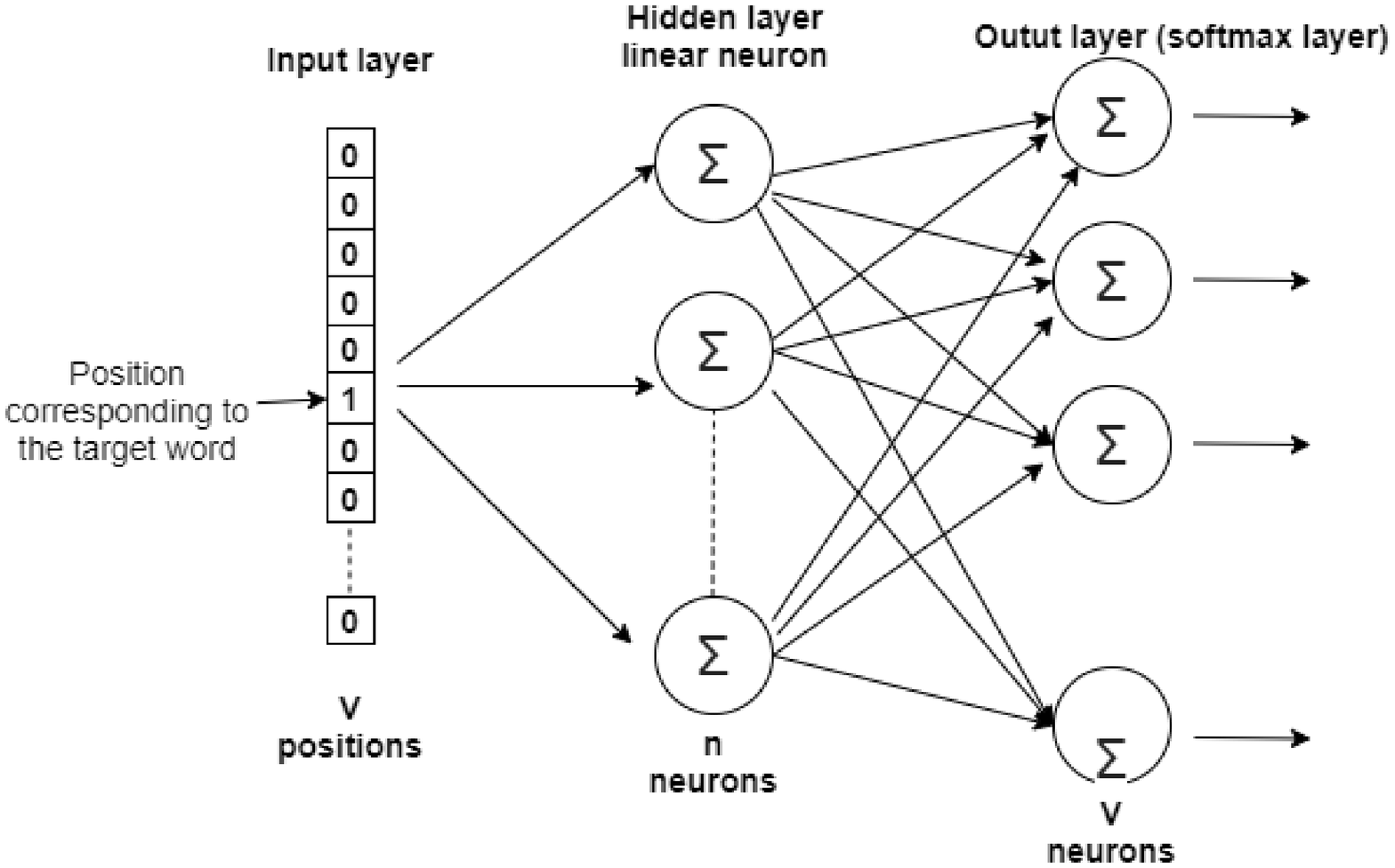}
\caption{Graphical representation of Skip-gram model. \cite{44} }
\label{Figure:figure3}
\end{figure}
The representation of skip-gram model is shown in the Figure \ref{Figure:figure3}. The number of neurons present in input and output layers are equal to the size of vocabulary $V$. The hidden layer consists $N$ number of neurons. Next, the one-hot representation  of target word is fed to the network. According to the one-hot representation, for a given target word $w_t$, only one of the $V$ positions, ($x_1, x_2, \cdots, x_V$) will be $1$, rest of the positions will be filled with $0$. For example, if the vocabulary size is $10$ and a word say, ``Big" is in the position $4$. Then, the one-hot representation of the word ``Big" is shown in the Table \ref{tab:table2}.
\begin{table}[!htbp]
\centering
\begin{tabular}{|l|l|l|l|l|l|l|l|l|l|}
\hline
\textbf{0} & \textbf{0} & \textbf{0} & \textbf{1} & \textbf{0} & \textbf{0} & \textbf{0} & \textbf{0} & \textbf{0} & \textbf{0} \\ \hline 
\end{tabular}
\caption{One-hot representation}
\label{tab:table2}
\end{table}
There is no activation function between input layer and hidden layer. Hence, the weighted sum of input is directly copied to hidden layer. The weight of the hidden layer is represented by the matrix say, $W_1$ with the dimension $V \times N$. Therefore, hidden layer (H) can be expressed as:
\begin{equation}
    H^T = W_1^{T} X={v^{T}_{w_t}}
\end{equation}
Now, assume there is $C$ context words. The output layer expecting $C$ multinomial distributions. Each output is computed using the weight matrix say, $W_2$ between hidden layer and output layer. The goal of this layer is to a set a parameter $\gamma$ such that it maximize the conditional probability $P(w_c \mid w_t)$ that means probability of the $w_c$ being predicted as the context of $w_t$ for all training pairs. If all the training pairs are denoted as $T$, this objective function can be expressed as;
\begin{equation}
   arg \: \underset{\gamma}{max}\prod_{(w_c, w_t) \in T}^{ } p(w_c \mid w_t, \gamma)
\end{equation}
In the skip-gram model, above conditional probability that is closeness of target word ($w_t$) and context word ($w_c$) is quantified using soft-max function as follows;
\begin{equation}
     p(w_c \mid w_t, \gamma) = \frac{e^{v_{w_c}}\cdot e^{v_{w_t}}}{\sum_{k=1}^{C}e^{v_{w_k}}\cdot e^{v_{w_t}}}
\end{equation}
where, $v_{w_c}$, $v_{w_t}$ $\in \mathbb{R}^m$ are vector representations of $w_c$ and $w_t$. $C$ is the all contexts. The parameters $\gamma$ are $v_{w_{c_i}}$, $v_{w_{t_i}}$ for $w \in V$, $c \in C$ and $i= [1,2, \cdots, m] $.

The above equation can be re-written by taking logarithm form and switching product to sum.
\begin{equation}
    arg\, \underset{\gamma}{max}\sum_{(w_c, w_t) \in T}^{}log (p(w_c \mid w_t)) = \sum_{(w_c,w_t)\in T}^{} log \frac{e^{v_{w_c}}\cdot e^{v_{w_t}}}{\sum_{k=1}^{C}e^{v_{w_k}}\cdot e^{v_{w_t}}}
\end{equation}

Now, all the dissimilar word in reference to context are isolated. Further, the topic is assigned to every word vector according to the topic of target word and stored in a matrix say, $M_2$. A snapshot of $M_2$ is shown in Figure \ref{Figure:figure12}.
\begin{figure}[!tbp]
\centering
\includegraphics[width=5.5in]{W2V_out.eps}
\caption{Word-topic distribution using LDA and Word2Vec. }
\label{Figure:figure12}
\end{figure}
\subsubsection{Word tokenization} Word tokenization means  splitting of large text into several words. In order to decide the topic of papers, each title of the database $D$ is tokenized into words. Further, to make it more understandable, common words are removed from each title.
\subsubsection{Deciding dominant topic of papers} The goal of this last step is to decide the dominant topic of papers. A paper may contain multiple topics due to it's word distribution among several topics. Therefore, it is necessary to find most promising topic of a paper for categorisation. In this process, the database $D$ with tokenized words are employed. For a single word say, $w_i$, $i=[1,2,\cdots,n]$ belongs to each title, the corresponding topic is searched from improvised topic-word distribution matrix $M_2$. For each word ($w_i$), the topic is selected according to the topic of the word $w_i$ assigned in $M_2$. It may be noted that a word $w_i$ may belong to more than one topics. In this scenario, a specific topic for the word $w_i$ is selected based on the highest probability score to its topic in the word-topic distribution matrix $M_2$. Finally, the topic of a paper is finalized considering frequency of words belong to a topic. The overview of the approach is shown in the Algorithm \ref{algo:algorithm 1}.

\begin{algorithm}[H]
\DontPrintSemicolon
\SetAlgoLined
\KwIn{A Dataset consisting titles of $n$ papers, a matrix $M_2$ with topic-word distribution.}
\KwOut{Dominant topic of papers}
Tokenize each title $T_j$ into words.\;
Remove parts of speech.\;
\For{each title $T_j$}{
\For{each word $w_i \in T_j$}{
Go to $M_2$\;
Search all assigned topics for the word $w_i$\;
Note the corresponding probabilities.\;
Select the topic with high probability and assigned to $w_i$.
}
Count the frequency of each topic.\;
Set the highest frequency topic as a dominant topic of papers.\;
\lIf{two topics say, $t_1$ and $t_2$ exist with maximum frequency}{\;
Check the probabilities of words under these $t_1$ and $t_2$.\;
The higher probability is considered for selection of one topic.
}
\caption{{\sc Algorithm 1 (Dominant topic selection)}}
\label{algo:algorithm 1}
}
\end{algorithm}
\subsection{User intention prediction} 
\begin{figure}[!htbp]
\centering
\includegraphics[scale=0.5300, angle=0]{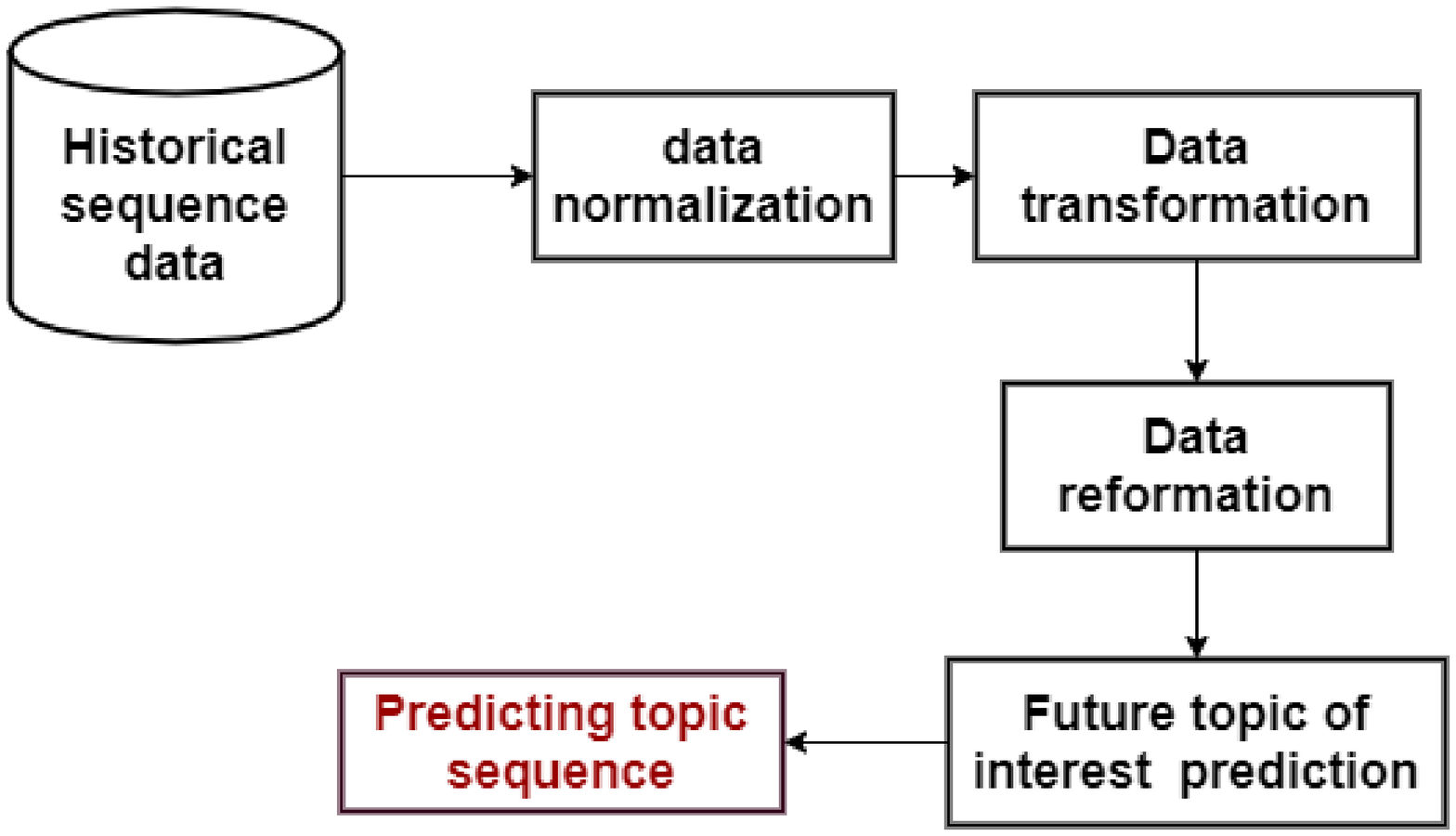}
\caption{Summary of user intention prediction. }
\label{Figure:figure4}
\end{figure}

Extracted preference which is obtained in the user preference extraction method discussed in Section $3.1$ can be used as a feature of user profile in research paper recommendation system. A user profile is categorized in two types: (1) static and (2) dynamic. Static profile contains such kind of user information (e.g. age, sex ) which does not require any modification. Generally, the information of the profile is supplied by the user himself. In contrast, dynamic profile is automatically generated by the recommendation system and the features which it contains, undergo changes over time. Since, the proposed approach captures the topic of interest of a user which may change in different context or time, it is dynamic. It will update accordingly as well as  increase in size and variation. In future, a range of variations in topic preference will increase the difficulty of decision task in recommendation. For example, if a user profile contains $(t_1, t_4, t_6, t_1, t_7)$ as preferable topics, it is difficult to predict that which topic of papers the user will want in next session. To mitigate this problem, it is required to analysis all the historical interactions of a user. In this regard, the conventional approaches of modelling user profile struggles to get all historical sequences(long term and short term) of user-item interaction \cite{33}, hence, leads to imperfect user modelling.  In this scenario, sequential modelling is good choice in academic and practical application. However, in sequential model time plays an important role. Also, time differentiates the user's interest in short-term and long-term categories. The short-term interest reflects the current interest of a user, which is changeable. In opposite, long term interest are more stable. The sequential modeling efficiently capture user's long term preference across different sessions as well as short term preference within a session. In case of sequential modeling, deep learning based methods has gained a lot of attention than machine learning based method such as Markov chain \cite{34}, session based k-NN \cite{35}. The proposed approach considers a deep sequential topic analysis technique to predict the future topic of interest of a user from his past preferable sequence of topics. Specifically, a variation of recurrent neural network (RNN) that is Long Short Term Memory (LSTM) is used to combine long term and short term interest about topic to predict the future topic of interest. 
Let, the historical topic sequence of a user $u_i$ is denoted by $H^u$ and defined by $H^u = [(({tp_1, tp_3})^u, {tm_1}^u), (({tp_2,tp_5})^u, {tm_2}^U), \cdots, (({tp_{j_1}, tp_{j_2}})^u, {tp_j}^u)]$, where $((t_{i_1},t_{i_2})^u,(tm_i)^u)$ means that a user ${u_i}$ likes ${i_1, i_2}$ topics at time $tm_i$. The task is to predict next preferable topics $tp_l \in tp$ given a certain user $u_p$ at a certain time $tm_q$. In addition, the proposed approach considers temporal features such as ``time difference "  between two clicked papers. In addition, few external features such as ``Liked" (whether the user liked the paper or not), ``session number" is taken into account. The overview of the proposed approach is shown in Figure \ref{Figure:figure4}.
\subsubsection{Data normalization} Normalization is required to scale data from it's original range to a range of values 0 to 1. Normalization helps the prediction model to learn optimal parameter easily of each input. The normalization of a value $x$ can be executed using following equation.
\begin{equation}
    x = \frac{(x - min)}{(max - min)}
\end{equation}

\subsubsection{Data transformation}The next task is to transform the sequential data into supervised data, since neural network is a supervised model. Therefore, the data should be in the format like $feature \rightarrow target$. In keras \cite{36}, there is a function ``look back", which efficiently transform sequential data to supervised data. Look-back is used to process past data upto ($t - lookback$) to predict at time $t$. For example, if the sequence data is like $[1 \; 2 \; 3 \; 4 \; 5]$ and look-back is $2$, the transformed data will be looked as follows;
\begin{equation}
 \begin{matrix}
[1 \; 2] & \rightarrow & [3] \\ 
[2 \; 3] & \rightarrow & [4] \\ 
[3 \; 4] & \rightarrow & [5]
\end{matrix}
\end{equation}

\subsubsection{Data reformation} One of the important fact about the LSTM model is the 3-dimensional input format of data. Therefore, it is required to reshape the 2-dimensional data into 3-dimensional form such as (batch\_size, time steps, input\_dim). In this context, the number of time steps is equal to number of LSTM cells, input\_dim is equal to number of features, and batch\_size is the number of windows of data that has to be passed at once.

\subsubsection{Future topic of interest prediction}
\begin{figure}[!htbp]
\centering
\includegraphics[scale=0.42000, angle=0]{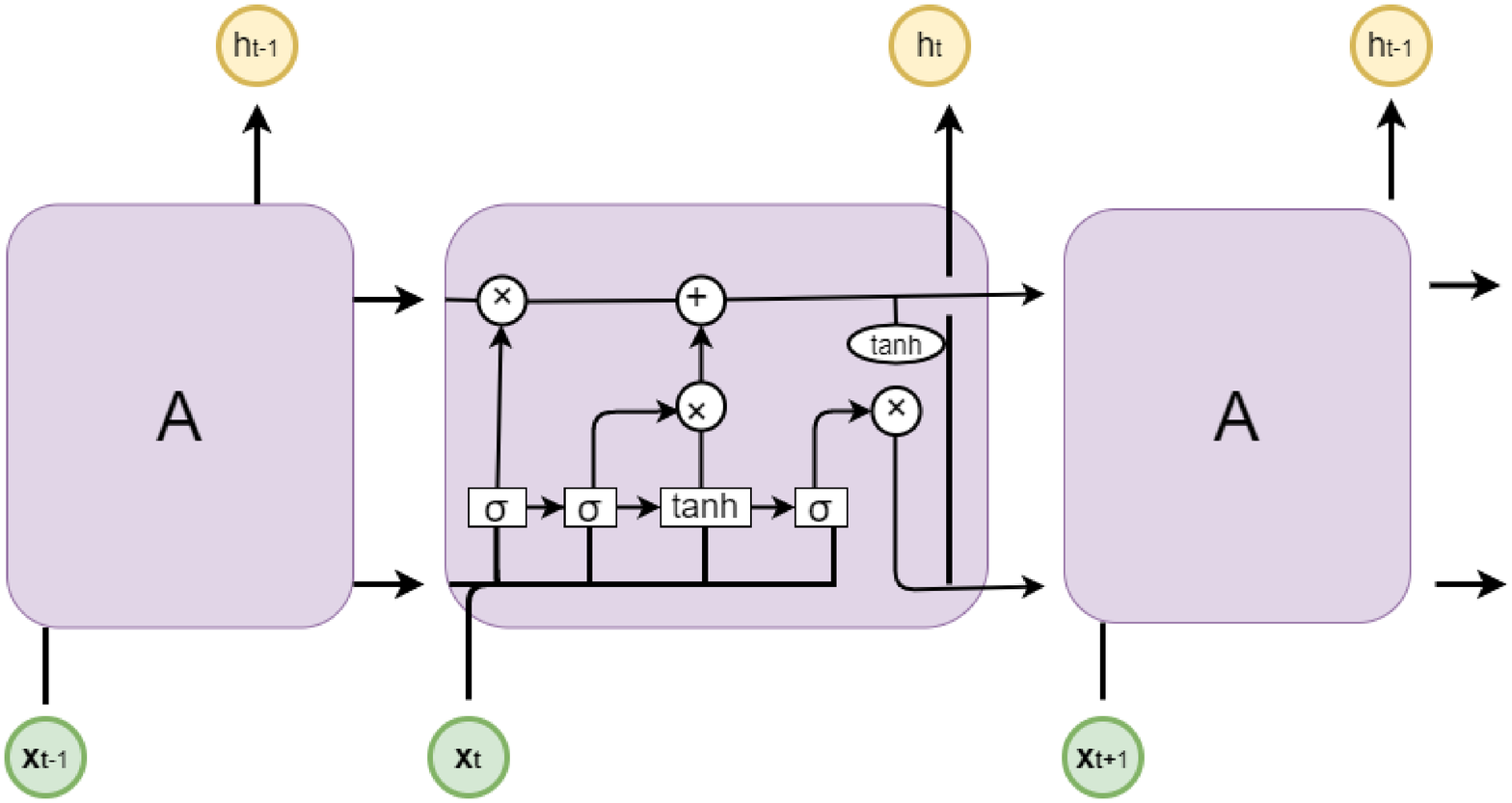}
\caption{LSTM architecture of the topic of interest prediction. \cite{43}}
\label{Figure:figure5}
\end{figure}

After the reformation of data, the immediate task is to generate a LSTM model to train the data. The architecture of LSTM is shown in the Figure \ref{Figure:figure5}. 
It has been seen from the Figure \ref{Figure:figure4}, LSTM is a sequential model, where several neural network module are connected sequentially. Each LSTM cell consists four gates with a common cell state to control retention and updating information learned from sequence data. The description of gates and cell state as follows;
\begin{itemize}
    \item input gate : It consists input vector.
    \item forget gate : It decides the amount of information to be allowed.
    \item output gate : It consists output vector generated by each LSTM cell.
    \item cell state : It runs through the entire network and carries information. LSTM has the ability to add or remove information to cell state using gates.
\end{itemize} 
Let, $x_t= [x_1, x_2, \cdots, x_n]$ is the input vector at time $t$, $h_t$ is the output vector at time $t$, and information of cell state at time $t$ is denoted by $c_t$. The first step of the LSTM is to decide which information of the cell state $
c_{t-1}$ has to be removed. The decision has been taken by the forget gate using a sigmoid function ($\sigma$). The forget gate uses the input values $x_t$ and output values at previous time steps i.e.$h_{t-1}$ to decide the output between 0 and 1. $0$ indicates the completely remove of the value. On the other side, $1$ indicates to keep the value completely. The mathematical expression of this function is;
\begin{equation}
    f_t = \sigma(V_f \cdot [h_{t-1},x_t]+B_f)
\end{equation}
Where, $V_f$ and $B_f$ are the weight and bias at forget layer.

Further, the cell state stores new information executing two steps.  First, the input gate choose which values will be updated using sigmoid function. Second, a $tanh$ function is used to create a new vector $\widehat{c_t}$ to add to the previous states values. Mathematically, the above two steps are expressed as;
\begin{equation}
    i_t = \sigma (V_i \cdot [h_{t-1},x_t]) + B_i)
\end{equation}
where, $V_i$ and $B_i$ are weight and bias at the input gate layer.
\begin{equation}
    \widehat{c_t} = tanh(V_c \cdot [h_{t-1}, x_t] + B_c)
\end{equation}
where, $V_c$ and $B_c$ are weight and bias simultaneously.
Finally, these two values are combined to update the state values as follows;
\begin{equation}
    c_t = f_t \odot c_{t-1} + i_t \odot \widehat{c_t}
\end{equation}
The cell state is updated adding new value with the values selected by forget layer at previous time step. At the end, the output is decided by the output gate after filtering with sigmoid function ($\sigma$). After that, the filtered vector is multiplied by $tanh$ function to get the output in the range between $-1$ to $1$. The execution at output layer can be expressed with the following two equations.
\begin{eqnarray} \label{eqn:15}
    o_t & = & \sigma(V_o \cdot [h_{t-1},x_t] + B_o)
    \nonumber\\
    h_t & = & o_t \times \tanh{c_t}
\end{eqnarray}
The equation \ref{eqn:15} determines the portion of the current state is allowed to be shown as output and can be used for next iteration of training.

The parameter of the LSTM used in this approach is optimized by ``Adam" , a variation of Stochastic Gradient Descent optimizer. Adaptive Moment Estimation (Adam) computes adaptive learning rate for each parameter efficiently and very fast.

\section{Experiment and experimental result}\label{sec:exp} This section presents objectives of the experiment, dataset description, experimental set up, procedure, evaluation metrics and results observed.

\subsection{Objectives of the experiments} The objectives of this experiment are finalized to answer the following research questions:

\begin{enumerate}
    \item \textbf{RQ1}: Does the proposed hybrid topic extraction model is comparable to that of the  state of the art topic models?
    \item \textbf{RQ2}: How much effective is the LSTM-based user intention model?
    \item \textbf{RQ3}: How does the proposed approach influences the performance of a research paper recommendation system?
\end{enumerate}

\subsection{Data set}The proposed approach considered two data set. The first data set includes corpus of titles was collected from Scopus \footnote{DataSources: \url{https://www.scopus.com/search/form.uri?display=basic}}. This data set was used to generate topic of papers using the proposed hybrid topic extraction model. In this work, titles of $2000$ papers were considered. Second, user's preference data is used to evaluate proposed sequential model. This preference data was collected from 12/10/2019 to 18/5/2020 using the proposed recommendation system. The statistics of the preference data set are shown in Table \ref{tab:table3}.
\begin{table}[!htbp]
\centering
\caption{statistics of data set}
\label{tab:table3}

\begin{tabular}{|l|l|}
\hline
\multicolumn{2}{|c|}{Data set 2} \\ \hline
Number of users       & 50     \\ \hline
Number of items       & 5213   \\ \hline
Number of features    & 3      \\ \hline
Name of features & \begin{tabular}[c]{@{}l@{}}Topic,Time difference, \\ Session no\end{tabular} \\ \hline
\end{tabular}
\end{table}

\subsection{Experimental environment} The proposed approach is implemented in Google Colab Notebook 4. All the codes are implemented and executed Python version 3.6 and keras programming environment.

\subsection{Experimental procedure}
Experiments were started with the preparation of data set, which was as follow.


\indent\textbf{Data preparation: } For the preference learning model, a data set was prepared comprising research paper titles. Further, for evaluating a user's intention prediction model, the user's preference data were collected from a user's log data maintained in the system's database. From the user's log data, clicked papers were considered. Next, all the required features such as titles, topics, time differences, session numbers were extracted as the user profile.  

\subsection{Experiments vis-a-vis objectives}
\indent \textbf{Experiment 1} To show proposed hybrid topic extraction model performs better than LDA alone (RQ1).

\begin{figure}[!htbp]
\centering
\includegraphics[scale=0.54000, angle=0]{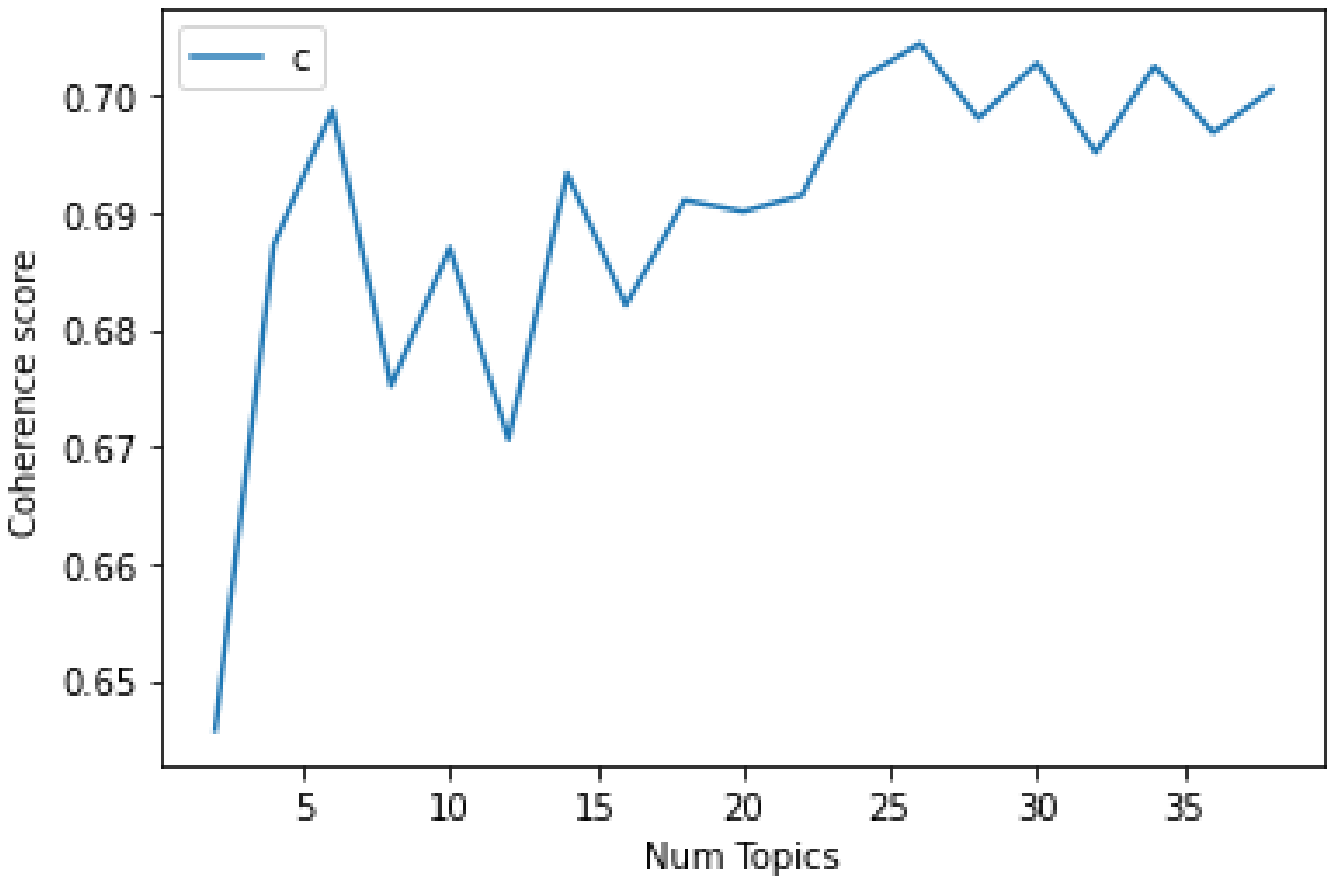}
\caption{Topic coherence score. }
\label{Figure:figure6}
\end{figure}

\begin{figure}[!htbp]
\centering
\includegraphics[scale=0.42000, angle=0]{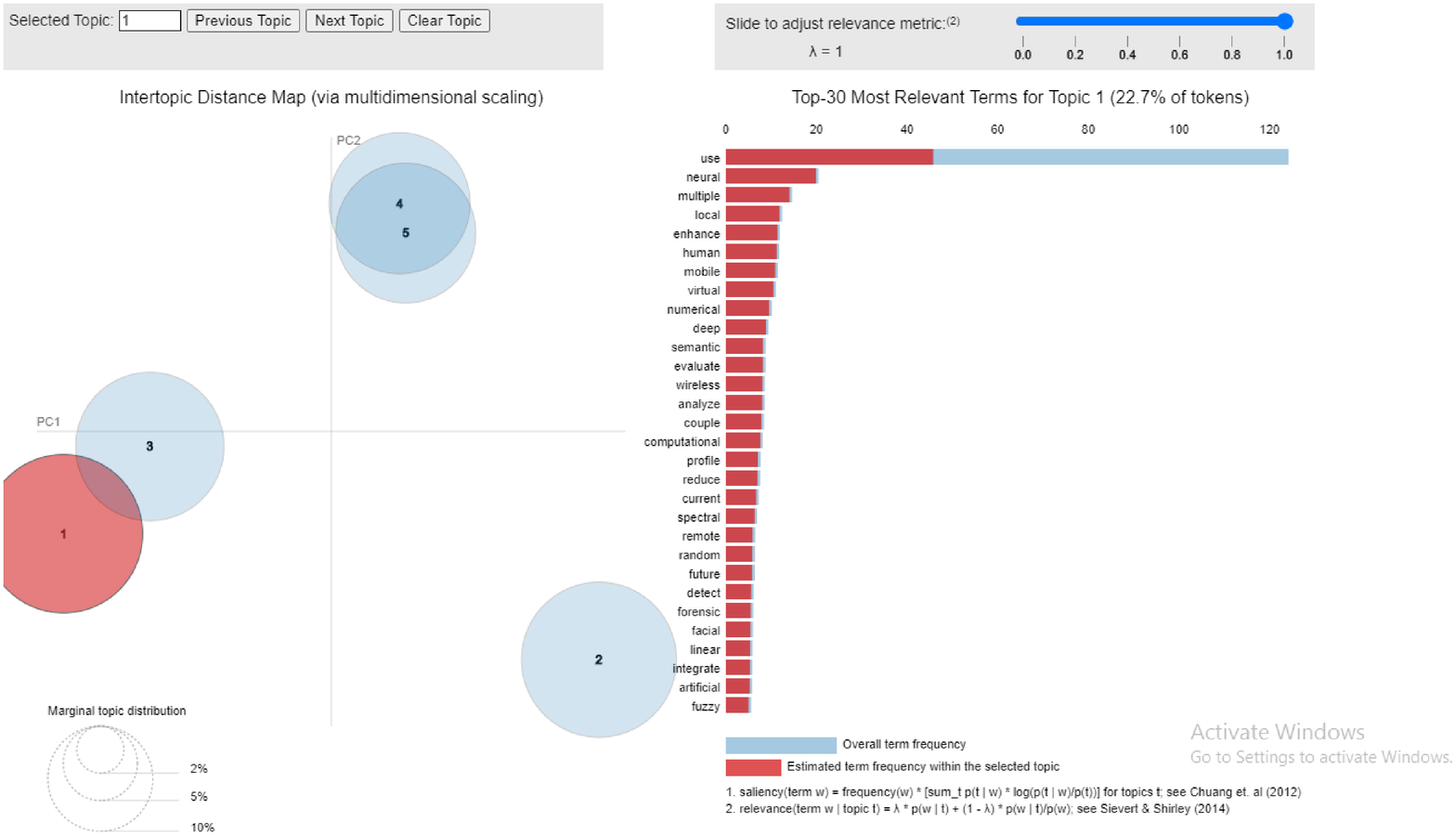}
\caption{Result of LDA for topic 1. }
\label{Figure:figure7}
\end{figure}

\indent The goal of this experiment to prove the efficacy of the proposed hybrid topic extraction model. To show the better performance of the proposed model over LDA alone, The experiment was divided into two parts. In the firs part, only LDA was applied on the dataset1 and decided topics of papers and add it to dataset1. The number of topics are decided using topic coherence score. the result is shown in the Figure \ref{Figure:figure6}. Further, Figure \ref{Figure:figure7} represents the distribution of words for a specific topic. Next, the dataset1 was split into training and testing set. Further three classifiers such as Support Vector Machine (SVM), Logistic Regression (LR), and Random Forest (RF) were applied to training data and validated on the test data. The model was validated using average value of F1 micro and F1 macro obtained from 5-fold cross validation. F1 micro and F1 macro are broadly used to validate classifier on multi class data \cite{42}. The results are shown in Table \ref{tab:table5}.
 
\begin{figure}[!htbp]
\centering
\includegraphics[scale=0.45000, angle=0]{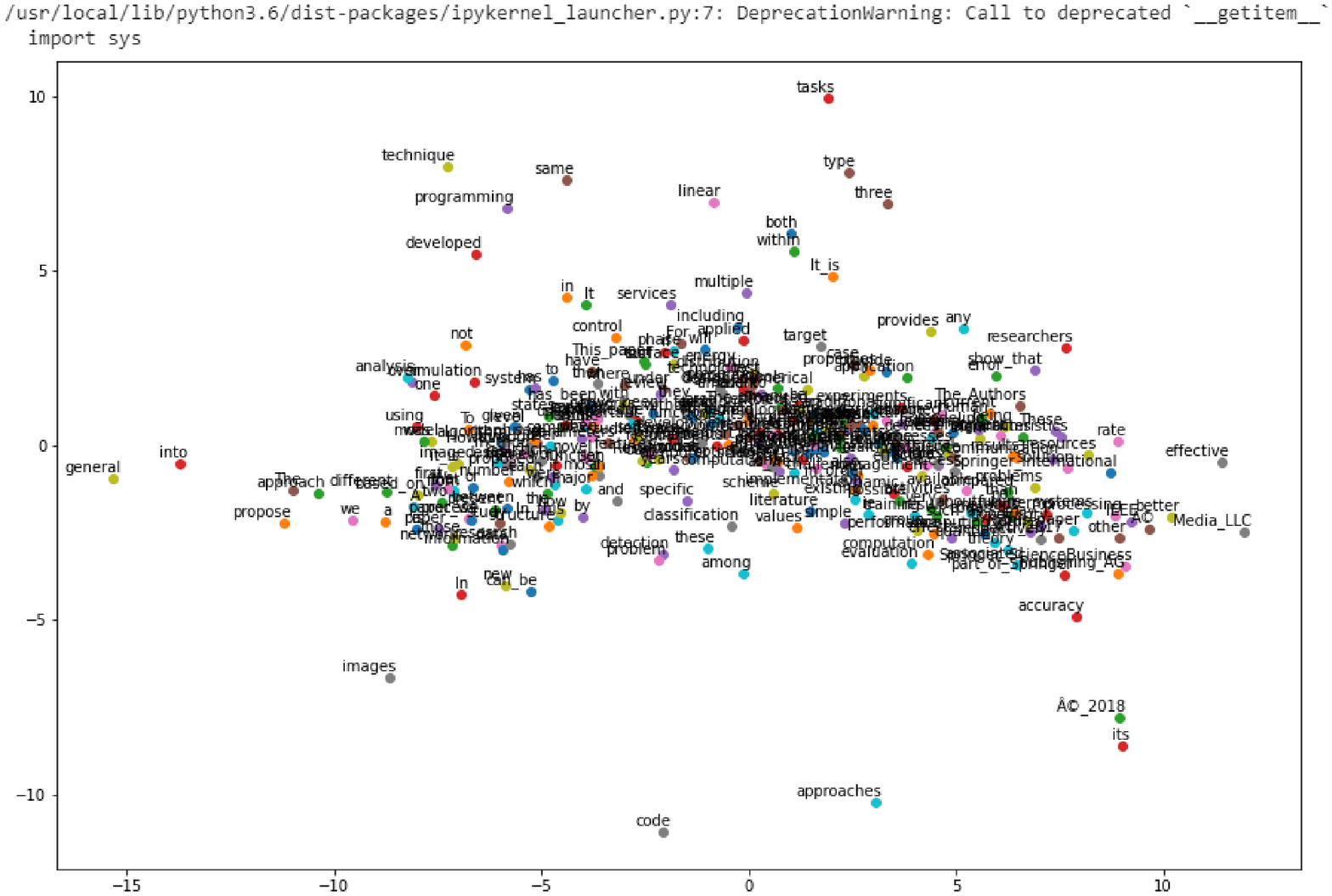}
\caption{Visualization of closed words. }
\label{Figure:figure8}
\end{figure}

In the second part, proposed hybrid model was applied on the same dataset and assigned topics of papers to dataset1 according to that. The word2vec was trained with the own dataset. For the training, the parameter of word2vec were set as like, window size was 5, dimension was 200, and, min count 5. The model was implemented using gensim word2vec model. In particular, skip-gram, a variation of word2vec was implemented in keras. t-SNE was used to visualize the words across models. Figure \ref{Figure:figure8} shows the vocabulary of words generated from word2vec. Finally, on the resultant dataset, three classifiers were applied likewise experiment 1. Table \ref{tab:table5} presents the classifier's result in terms of F1 micro and F1 macro.

\begin{table}[!ht]
\caption{Classification results.}
\label{tab:table5}
\resizebox{12cm}{!}{
\begin{tabular}{|l|l|l|l|l|l|l|}
\hline
\textbf{Topic model} & \multicolumn{2}{c|}{\textbf{SVM}} & \multicolumn{2}{c|}{\textbf{LR}} & \multicolumn{2}{c|}{\textbf{RF}} \\ \hline
               & F1 micro & F1 macro & F1 micro & F1 macro & F1 micro & F1 macro \\ \hline
LDA            & 55.70\%  & 54.44\%  & 51.32\%  & 50.67\%  & 54.39\%  & 52.77\%  \\ \hline
LDA + Word2Vec & 72.13\%  & 74.48\%   & 62.6\%   & 64.13\%  & 64.55\%  & 66.12\%  \\ \hline
\end{tabular}}
\end{table}

From Table \ref{tab:table5}, it has been proved that proposed hybrid model performs better than LDA topic model for all cases.

\indent \textbf{Experiment 2} To show the sequential model considered in this work performs better than the existing models (RQ2).

\begin{figure}[!htbp]
\centering
\includegraphics[width=4.8in]{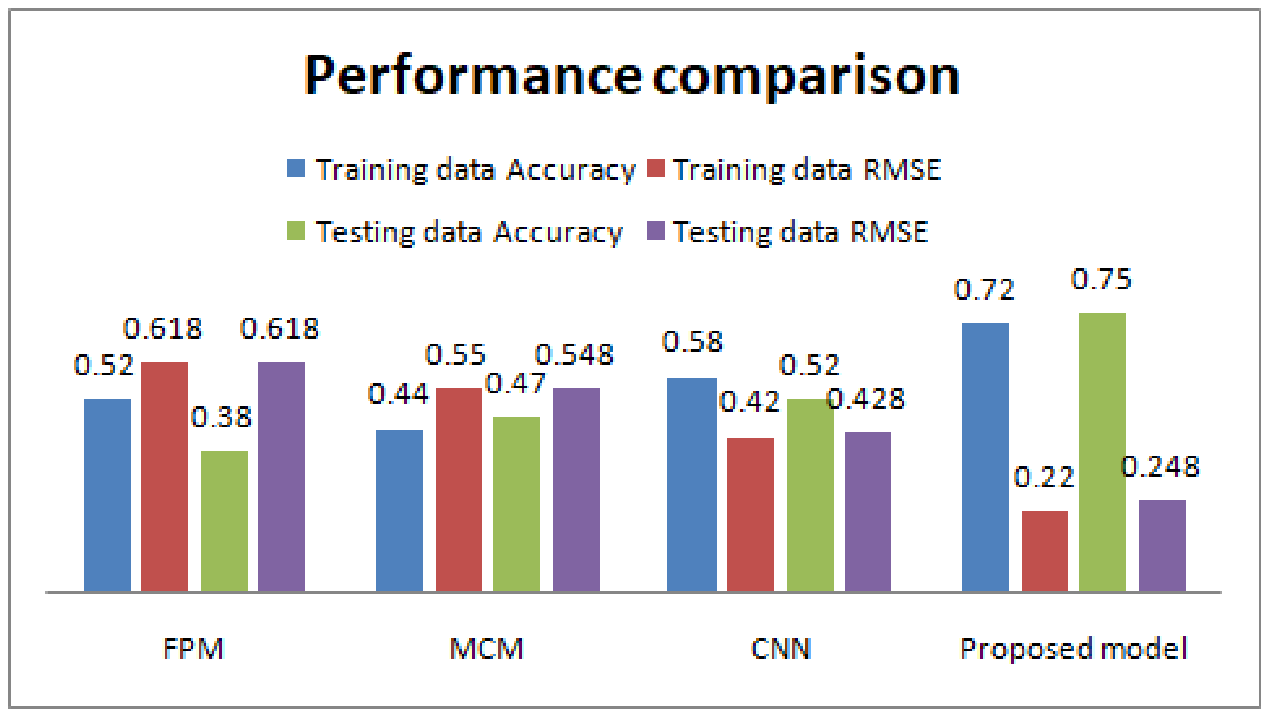}
\caption{Performance comparison with the existing models. }
\label{Figure:figure9}
\end{figure}

\indent This experiment was conducted considering the dataset 2 that is historical interaction data of users and divided into two parts training and testing data. Further, LSTM based sequential model was applied on the training and testing data successively. Finally, the performance of the model were decided using the metric Accuracy and Root Mean Square Error (RMSE) over both the training and testing data. After that, this performance was compared with three existing sequential models such as Frequent pattern mining (FPM), Markov Chain Model (MCM), Convolution Neural Network (CNN). In this experiment, same dataset was considered for evaluating all models. The existing models are described as follows:
\begin{itemize}
\item \textbf{FPM: } Frequent Pattern Mining (FPM) is a data mining technique. FPM was used on historical sequential topic data to predict future topic of interest.
\item \textbf{MCM: } Markov Chain Model (MCM) was applied on the user preference data and predict future preference.
\item \textbf{CNN} Convolution Neural Network (CNN) is a deep learning model. CNN was applied in similar way to predict future topic preference.
\end{itemize}
Table \ref{tab:table4} and Figure \ref{Figure:figure9} presents the results of performance metrics. From the results, it has been observed that the sequential model, used in this work, performed better on both train and test dataset.
\begin{table}[!ht]
\centering
\caption{performance comparison with existing model}
\label{tab:table4}
\begin{tabular}{|c|c|c|c|l|}
\hline
        & \multicolumn{2}{c|}{Training data}   & \multicolumn{2}{c|}{Testing data} \\ \hline
Metric  & Accuracy & \multicolumn{1}{l|}{RMSE} & Accuracy          & RMSE          \\ \hline
FPM     & 0.52     & 0.618                     & 0.38              & 0.618         \\ \hline
MCM     & 0.44     & 0.55                      & 0.47              & 0.548         \\ \hline
CNN     & 0.58     & 0.42                      & 0.52              & 0.428         \\ \hline
LSTM & 0.72     & \multicolumn{1}{l|}{0.22} & 0.75              & 0.248         \\ \hline
\end{tabular}
\end{table}

\indent \textbf{Experiment 3} To compare the result of the paper recommendation according to user evaluation with the results obtained from existing search engine (RQ3).

\indent In order to prove the significance of the proposed approach based on user perspective, a user survey was conducted. In this survey, 100 users were selected from researcher group. 

\indent For each participant, 10 sessions was conducted. In each session, four baseline models and the proposed model were used arbitrarily and users were unaware about the type of models. Before starting the experiment, each participant was requested to give 10 keywords from their area of interest. From these 1000 keywords, collected from all participants, duplicates keywords were removed. Finally, 640 keywords were selected randomly and divided into 40 sets. In each session participant choose a set randomly. In the first session, candidate papers that against the search query  were extracted from the dataset and provided to the participant. Then, the response from them were recorded. Further, all the required features were extracted from their clicked papers and stored in a user profile. From next time onward, this profile was used to predict the user intention and recommend to that specific participant. After every session, profile was updated. Finally, the average results of nine sessions were considered as a final evaluation of the users.

\indent To compare the performance of the proposed model three popular search engine, namely, Google Scholar (GS), Microsoft Academic Search
(MAS), and (3) Citeseer were considered. The experiment 2 was repeated for evaluating the results of search engines.   

\indent The following four metrics such as Recall@10, Precision@10, MAP@10 and CTR were used to evaluate the results. All metrics were computed based upon the judgment of users. The definition of above metrics are as follows:
\begin{enumerate}
    \item Recall@k: Recall@k can be defined using the following equation;
\begin{equation}
    Recall@10 =\frac{P_{rec} \cap P_{rev}}{P_{rev}}
\end{equation}
$P_{rec}$ is the number of recommended papers and $P_{rev}$ is the number of relevant papers according to users. In this context, user's clicked papers were assumed as relevant papers. 
    \item precision@10: Precision@10 can be defined as follows;
\begin{equation}
    Precision@10 =\frac{P_{rec} \cap P_{rev}}{P_{rec}}
\end{equation}
    \item MAP@10: This is the average of reciprocal ranks of a paper in the recommendation list. The reciprocal rank is set to 0 if the rank is above 10. MRR@10 takes into account the rank of the item. Each metric is evaluated 10 times and averaged.
    \item CTR rate: It measures how many recommended papers are clicked by the user.
\end{enumerate}

\begin{figure}[!htbp]
\centering
\includegraphics[width=5.0in]{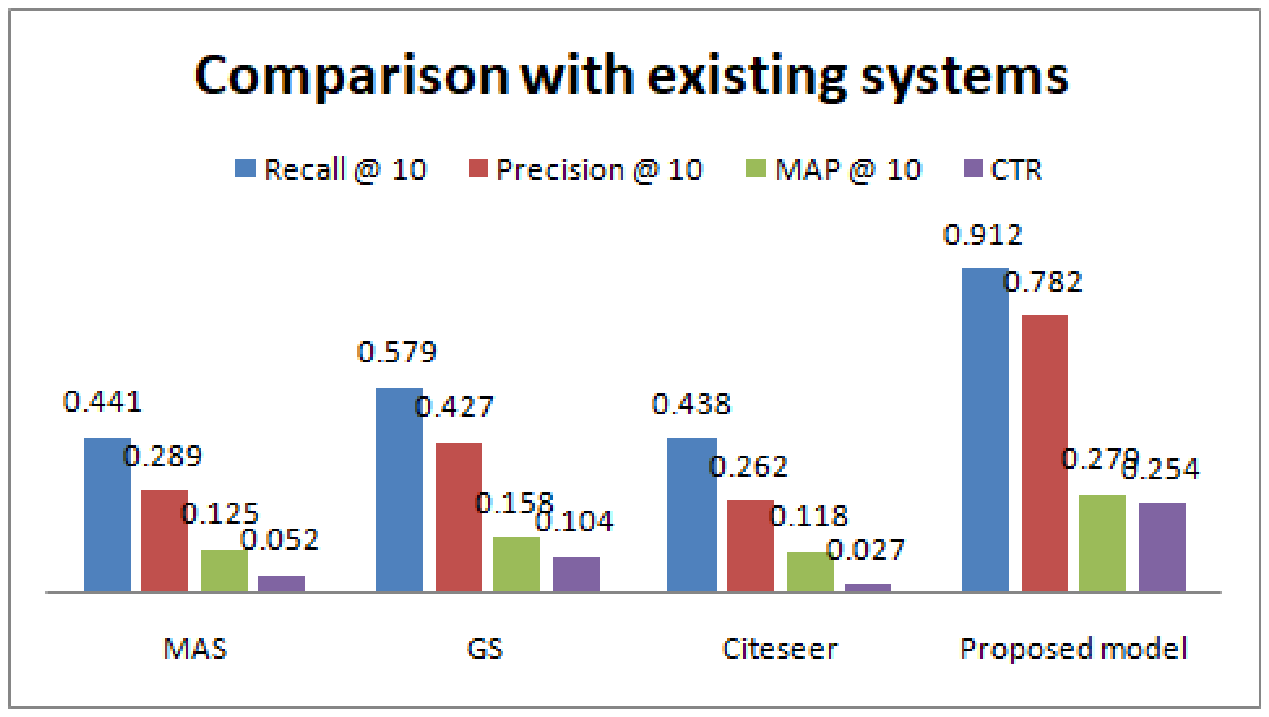}
\caption{User opinion based comparison }
\label{Figure:figure10}
\end{figure}

\indent\textbf{Result} The comparative result is shown in Figure \ref{Figure:figure10} . From the results, it is observed that the model comprising proposed method outperforms the search engines.

\section{Validity limitations} \label{sec:tv}The claims of this work are under some assumptions and limited considerations. This section highlights the considered features and other ways of improvement.

\begin{itemize}
\item \textbf{Times tamp: } This work utilized an LSTM model to predict a user's intention where time stamps were used as the contextual feature. However, it could not properly capture the drift of a user's intention. It can be modeled better using timeshare LSTM. 

\item \textbf{External context: } Other than time, other external contexts, such as user experience, user understanding level, author name, etc. can be used to enhance the prediction.

\item \textbf{Item-item relation:} Item-item relation is also an important factor to understand the user intention, which is not incorporated in the proposed approach.
\end{itemize}

\section{Conclusions}
\label{sec:con} This work aims on modeling user intention needed for  a research paper recommendation system. The first step is to categorize a paper into a topic. A hybrid topic model (HTM) comprising LDA (Latent Dirichlet Allocation) and Word2Vec is proposed. HTM decides the topic of a paper considering probability distribution of words among several topics and correlation among words. The result establishes that HTM  performs better categorization than either LDA or Word2Vec models.  In the second step, a sequential LSTM model is employed to predict the intention of a user in terms of her topics of interest. The considered LSTM model predicts the intention of the user from the history of interaction  and time interval as a contextual feature. This essentially captures long term and short term interests of users and thus dynamic nature of users'  intentions. The experimental results substantiates the efficacy of the proposed approach.   In future, this work can be extended to explore more features as users' preference and a recommendation technique using multi-criteria preference analysis.

\bibliographystyle{spphys}       
\bibliography{ref}   

%
%

\end{document}